\documentclass[sigconf]{acmart}
\usepackage{todonotes}
\usepackage{subcaption}
\usepackage{booktabs}
\usepackage{colortbl}
\usepackage{paralist}
\usepackage{stfloats}
\AtBeginDocument{%
  \providecommand\BibTeX{{%
    \normalfont B\kern-0.5em{\scshape i\kern-0.25em b}\kern-0.8em\TeX}}}

\acmYear{2022}
\acmConference[WPES '22]{Proceedings of the 21st Workshop on Privacy in the Electronic Society}{November 7, 2022}{Los Angeles, CA, USA}
\acmBooktitle{Proceedings of the 21st Workshop on Privacy in the Electronic Society (WPES '22), November 7, 2022, Los Angeles, CA, USA}
\acmPrice{}
\acmDOI{10.1145/3559613.3563205}
\acmISBN{978-1-4503-9873-2/22/11}

\begin{document}

\title{Your Consent Is Worth 75 Euros A Year -- Measurement and Lawfulness of Cookie Paywalls}

\ccsdesc[500]{Security and privacy~Economics of security and privacy}
\ccsdesc[500]{Information systems~Online advertising}

\author{Victor Morel}
\affiliation{%
  \institution{Vienna University of Economics and Business}
  \city{Vienna}
  \country{Austria}}
  \email{contact@victor-morel.net}

\author{Cristiana Santos}
\affiliation{%
  \institution{Universiteit Utrecht}
  \city{Utrecht}
  \country{Netherlands}}
\email{c.teixeirasantos@uu.nl}

\author{Yvonne Lintao}
\affiliation{%
  \institution{Vienna University of Economics and Business}
  \city{Vienna}
  \country{Austria}}
\email{yvonne.lintao@s.wu.ac.at}

\author{Soheil Human}
\affiliation{%
  \institution{Vienna University of Economics and Business}
  \city{Vienna}
  \country{Austria}}
\email{soheil.human@wu.ac.at}

\renewcommand{\shortauthors}{Morel \textit{et al.}}

\begin{abstract}
Most websites  offer their content for  free, though this  gratuity often comes with a counterpart: personal data  is collected  to finance these websites by resorting, mostly, to tracking and thus targeted advertising.
Cookie walls and paywalls, used to retrieve consent, recently generated interest from EU DPAs and seemed to have grown in popularity.
However, they have been overlooked by scholars.
We present in this paper 1) the results of an exploratory study conducted on 2800 Central European websites to measure the presence and practices of \textit{cookie paywalls}, and 2) a framing of their lawfulness amidst the variety of legal decisions and guidelines.

\end{abstract}

\keywords{paywalls,  tracking, dark patterns, adtech, consent, GDPR, ePD}
\maketitle

\section{Introduction}

Most websites  offer their content for  free, though this  gratuity often comes with a counterpart: the  personal data  of users is collected  to finance these websites services by resorting, mostly, to  tracking and thus targeted advertising.
 Websites may use \textbf{cookie walls}, also known as tracking walls or \textit{"take it or leave it choices"}~\cite{j_zuiderveen_borgesius_tracking_2017, edps_edps_2017} 
which condition the website access to  the  user's acceptance of certain trackers on her device (computer or smartphone)~\cite{Sant-etal-20-TechReg},
 whereas data subjects are more likely to consent if given such take-it-or-leave options, even if they do not want to provide their personal data in exchange for visiting a website~\cite{j_zuiderveen_borgesius_tracking_2017}.
Websites also developed %
\textbf{paywalls} to enforce their subscription business models: a user who refuses to accept tracking is then obliged to provide a sum of money to access that website~\cite{cnil_cookie_22}.
A paywall thus consists of \textit{“various strategies that restrict content access until the user pay for them, possibly after allowing them to view some content for free”}~\cite{papadopoulos_keeping_2020}.

Paywalls have been questioned from a legal point of view by several non-academic stakeholders~\cite{noyb_news_2021}\cite{cnil_cookie_22}
because consent to the processing of personal data must be given freely and unconditionally (Art. 7(4) GDPR), and consent is not "freely given" if users cannot refuse tracking without adverse consequences, \textit{e.g.}, by paying a subscription.
European Data Protection Authorities (DPAs)~\cite{cnil_cookie_22, datenschutz-behorde_faq_2022} %
recently  %
consider paywalls legitimate, %
though there is some inconsistency in the positions taken on whether a paywall is legally compliant with the General Data Protection Regulation (GDPR)~\cite{european_parliament_general_2016} and with the ePrivacy Directive (ePD)~\cite{european_parliament_directive_2002}.

Paywalls generally offer the possibility to give  access to the content of the website  \textit{without} tracking, against payment or an account  subscription, \textit{i.e.} a paid tracking-free option. 
But the option of paying the subscription model is not actually free of charge or tracking-free because users might \textit{pay with their data}, even though they initially paid for an ad-free and tracking-free version of the website. 
Papadopoulos \textit{et al.}~\cite{papadopoulos_keeping_2020} report that users who pay for a subscription do not usually get a tracking-free version of the website since  after purchasing a subscription, they face as much advertising and tracking-related resources as possible.

Simply put, paywalls present several novel problems and %
their cross analysis 
with cookie walls, coined \textbf{cookie paywalls} (an illustration of which can be found in Figure~\ref{fig:derstandard} in the Appendix), 
has not been analyzed yet.
 To address this research gap, we investigated the following research questions: (1) What are the different types of paywalls and cookie walls, and how do they articulate with each other? (2)  What is the scale of these practices? (3)  Are these practices lawful?
Pursuant to these questions, we conducted an interdisciplinary and empirical study on the top of 150 websites
 in Central Europe %
 which resulted in the following contributions:
\begin{inparaenum}
    \item %
    a fine grained %
    classification of paywalls: \textit{hard}, \textit{soft}, \textit{metered paywall}, \textit{registration wall}, \textit{cookie paywalls}, and \textit{cookie wall},  %
    \item a legal analysis of  both cookie walls and paywalls, %
    \item a quantitative measurement of these practices:  %
    61 of the 2800 websites studied use paywalls (2.72\%), and 13 cookie paywalls (0.66\%);
    \item most cookie paywalls consist of news websites, %
    \item they do not track visitors prior to interaction, and
    \item their prices are within average for newspaper subscription.
\end{inparaenum} 

\vspace{-2mm} 

\section{Background and Related Work}
\label{sec:background}

\textbf{Legal requirements for online tracking.}
To comply with the GDPR and the ePD, websites must obtain \emph{consent} %
from EU users when tracking their  behavior (Art. 5(3) ePD) for concrete purposes (such as targeting advertising). 
Some other purposes are exempted of consent, \textit{e.g}, functional or essential trackers (Recital 66 ePD).
The only way to assess with certainty whether consent is required is to analyse the \emph{purpose} of each tracer on a given website~\cite{EDPB-4-12, WP203, Foua-etal-20-IWPE}.
A  valid consent  must comply with several requirements: \textit{prior}, \textit{freely given}, \textit{specific}, \textit{informed}, \textit{unambiguous}, \textit{readable}, \textit{accessible}, and \textit{revocable} (Art. 4(11) and 7 GDPR)~\cite{Sant-etal-20-TechReg}.
Most relevant to this paper are the requirements of prior and freely given consent. 

\noindent
\textit{Prior consent.}
Before storing or reading the trackers on a user's terminal, websites need to ask  consent to users (Article 6(1) GDPR).

\noindent
\textit{Freely given consent.} The request for consent should imply a voluntary  choice of the user to accept or decline some or all trackers, taken in the absence of any kind of pressure %
on the user to persuade her to give consent~\cite{wp29_opinion_2011, wp29_working_2013, wp29_guidelines_2017}. 
Making the provision of website service or access thereto  conditional on the acceptance  of certain non-essential trackers \textit{can affect}, in certain cases, the freedom of choice~\cite{cnil_cookie_22}, and subsequently, the validity of consent~\cite{wp29_guidelines_2017}.

\textbf{Paywalls and cookie walls.} %
Few scholars studied paywalls mostly from a business or a digital media point of view~\cite{carson_behind_2015, fletcher_paying_2017, myllylahti_what_2017}, though
the work of Papadopoulos \textit{et al.} \cite{papadopoulos_another_2020} is relevant to our paper since it is the first study investigating the privacy impact of paywalls.
They classify paywalls into two categories: i) \textbf{hard paywalls} \textit{"require paid subscription before a user  can access content"}, and ii) \textbf{soft or metered paywalls}  \textit{"limit the number of articles a viewer can read before requiring a paid subscription"}. This latter category is used indistinguishably. 
They %
found that 7,6\% of websites world-wide used paywalls in 2019, and that most of them are news websites.
Another finding  is that subscribing to a paying offer does not significantly reduce tracking. 
Santos \textit{et al.}~\cite{Sant-etal-20-TechReg} commented in 2020 that most stakeholders and regulators  consider that a cookie wall violates a freely given consent. 
Outside of the academic scholarship, the Austrian NGO NOYB filed complaints against paywalls of seven major German and Austrian news websites  in  2021~\cite{noyb_news_2021}  considering that consent is not freely given  if the alternative to being tracked is to "pay 100 times the market price of the user's personal data".

\textbf{Consent management.} %
Matte \textit{et al.} \cite{matte_cookie_2019} 
measured the %
compliance of IAB Europe’s Transparency and Consent Framework (TCF)
is the main framework used by several actors on the web to manage consent.
They found that out of  28 257 crawled websites,  1426  implemented the TCF, and further found  
consent may be stored prior to the user's consent,  or against their will. %
We build upon prior work and further address
\begin{inparaenum}
    \item the state of current paywall types,
    \item the usage by the most visited websites from different countries,
    \item the amount of money requested, and
    \item how current positions from regulators %
shape  paywall practices. %
\end{inparaenum}

\vspace{-2mm} 

\section{Methodology}
\label{sec:study}

\textbf{Target Countries.}
We examined  top websites of 13 countries that are categorised by \cite{Jordan2005} as a part of ``Central Europe'': Austria,  Croatia, Czech Republic, Estonia, Germany, Hungary, Latvia, Lithuania, Luxembourg, Poland, Slovakia, Slovenia, and Switzerland, a highly interesting target
region for our study: 
\begin{inparaenum}
    \item they are in the same geographical region, though %
    they show diversity in their language, culture, and economic situation, 
    \item the countries include former USSR republics, former eastern bloc countries, former western bloc countries, and neutral countries, as well as EU and non-EU countries, 
    \item these countries have  various DPA positioning regarding cookie paywalls (in favor for Austria, critical for Germany, without official statement for some others).
\end{inparaenum}
We aimed to investigate whether DPA positions is correlated with paywall practices.

\textbf{Data collection.} A list was constituted using Tranco  \cite{pochat_tranco_2019} and includes the top 1000 Austrian websites and the top 150 websites of each of the 12 other Central European countries (for a total of 2800) between 27th March and 25th April 2022. The  reason for conducting a more representative analysis of  Austrian websites relied on the fact that the Austrian DPA is the only one ruling in favour of paywall implementation~\cite{datenschutz-behorde_faq_2022} (see Table~\ref{table:requirements}). We aimed to study whether such a decision impacted the rate or type of paywalls used in Austrian websites. For statistical reasons, the results  restricted to the top 150 Austrian websites are presented distinctively (in parenthesis).

\textbf{Annotation and analysis.} 
Websites were inspected by 
three female students annotators  
which are native German speakers, and fluent in English. 
All non-German or non-English websites were translated using \url{deepl.com}. %
The annotation occurred between 19th May and 31th May 2022.
The annotators used Firefox 102.0.1 (64-bit) on the OS Microsoft Windows, free of add-ons and with \textit{Standard} privacy settings
unchanged after installation. 
The following aspects were  annotated:
i) whether the website has a cookie wall or paywall, 
ii) paywall type, 
iii) website category, and
iv) general notes (price
/type of subscription, position of the wall on the screen).
We initially based our paywall classification on~\cite{papadopoulos_another_2020} (explained in Section~\ref{sec:background}), but novel types of paywalls were identified (see Section~\ref{sec:results}.) %
The  analysis of websites consisted in two rounds of annotations.
The results were inscribed in a spreadsheet for each country. 
In rare cases of disagreements between the annotators, the annotators discussed the case verbally in online-calls until they reached a mutual agreement. 
The data was then merged into a single spreadsheet to allow the drawing of comparative analysis.

\textbf{Tracking assessment of cookie paywalls.}
Since  paywalls are claimed to propose an alternative to tracking, we analysed %
whether or not the websites implementing them were tracking their visitors before a user expresses their decision regarding payment \textit{by money} (one-time or standing payment) or \textit{with their data} (giving consent to tracking).
A fourth annotator with technical knowledge first visited the annotated websites listed as \textit{cookie paywalls} in the first round using Firefox 102 on Ubuntu 20.04 in incognito mode, with only the plugin Disconnect in monitoring and no-blocking mode.
We recorded the cookies
immediately after loading the page and without any previous user consent, and noted the number of trackers spotted by Disconnect.
In a second step, the annotator consented to all personal data collection and processing, refreshed the web-page and recorded the cookies and the number of trackers.

\textbf{Analysis of cookie walls and paywalls.}  The legal analysis is based  on the GDPR and the ePD.
We consulted the guidelines of both the European Data Protection Board ((EDPB -- an EU advisory board on data protection) and the guidelines and decisions of DPAs, in particular,  the very recent positions from 2022 of the French~\cite{cnil_cookie_22}, Austrian~\cite{datenschutz-behorde_faq_2022} and Italian DPAs~\cite{ItalianDPA-cookiewall-2021}.

\vspace{-3mm} 

\section{Results and discussion}
\label{sec:results}

\begin{table*}[!ht]
\scriptsize 
\caption{Paywall count per country. The numbers in parenthesis only consider the first 150 Austrian websites.}
\label{tab:results}
\hspace{-.7cm}
\begin{tabular}{@{}lllllllllllllll@{}}
\toprule
Total count & Austria & Croatia & Czech Republic & Estonia & Germany & Hungary & Latvia & Lithuania & Luxembourg & Poland & Slovakia & Slovenia & Switzerland & Total \\ \midrule
Hard cookie paywall & 5 (5) &  &  &  & 1 &  &  &  &  &  &  &  &  & 6 \\
Cookie wall & 6 (1) &  &  &  &  &  &  &  &  &  &  &  &  & 6 (1) \\
Soft paywall & 7 (4) & 2 & 3 & 3 & 1 &  & 4 & 1 & 2 & 1 & 3 & 1 & 6 & 34 (31) \\
Metered paywall &  & 1 &  &  &  &  &  &  & 1 &  &  &  &  & 2 \\
Hard paywall & 1 (1) &  &  & 2 &  &  &  &  &  &  & 1 & 1 & 1 & 6 \\
Cookie paywall + soft paywall &  &  &  &  & 7 &  &  &  &  &  &  &  &  & 7 \\
Registration wall &  &  &  &  &  &  &  &  &  &  &  &  & 1 & 1 \\
Total & 19 (11) & 3 & 3 & 5 & 9 & 0 & 4 & 1 & 3 & 1 & 4 & 2 & 7 & 61 (53) \\ \bottomrule
\end{tabular}
\end{table*}

\subsection{The results of the legal analysis}

\noindent
\textbf{Divergent positions on cookie walls and paywalls.} 
Borgesius \textit{et al.}~\cite{zuiderveen2017assessment}, in their EU commissioned study  on the Proposal for the ePrivacy Regulation, mentioned %
a non-exhaustive blacklist of circumstances in which cookie walls are 
banned:
\begin{inparaenum}[i)]
    \item tracking on websites, apps and or locations that reveal information about special categories of data,
    \item tracking by unidentified third parties for unspecified purposes,
    \item tracking by all government funded services,
    \item invalid consent (\textit{e.g.} unequal balance of power)
    \item consent asked for processing for multiple purposes.
\end{inparaenum}

The Austrian DPA~\cite{datenschutz-behorde_faq_2022} stated that, %
in principle, it is permissible to ask payment for access to a website as an alternative to consent to tracking. It reiterates his  position from 2018~\cite{AustrianDPA-cookiewall-2018} wherein it 
 indicated that paywalls give a degree of choice: i) accept tracking while accessing  a website; ii) refuse tracking with limited access to a website; or iii) pay subscription fees without  tracking.  %

The French DPA~\cite{cnil_cookie_22}
states that %
the lawfulness of  cookie walls and paywalls  must be assessed on a case-by-case basis, depending on a set of five evaluation criteria:

\noindent
\textit{Alternative to tracking}. 	When  users refuse  consent to tracking purposes,  websites need to show there is a "real and fair" alternative way (by other websites) allowing access to the content,  without consenting to data collection.
Possible imbalances need to be considered, %
for example if the website:
i)  has exclusivity on the content/service offered;
ii) is a dominant or essential service provider.

\noindent	
\textit{Reasonable price of a paywall}. 
Paywalls are  not prohibited, in principle, since it constitutes an alternative to consent to tracking. 
The paywall price must be reasonable. The threshold of the reasonable rate depends on a case-by-case analysis.

\noindent
\textit{Account requirements.} The %
creation of an account must pursue specific and transparent purposes for users (\textit{e.g.}, a subscription must be accessible on different terminals).

\noindent
\textit{Paywalls is limited to fair remuneration.} 
Paywall are limited to the purposes which allow fair remuneration for the service offered. If a publisher considers that the remuneration for its service depends on the income it could obtain from targeted advertising, only consent for this purpose should be necessary to access the service (refusing consent to, \textit{e.g.}, personalization of editorial content should not then prevent access to the content of the site).

\noindent	
\textit{Paywalls cannot  impose acceptance of all trackers in a website.} 
Websites may request the user's consent on a case-by-case basis to deposit trackers when the latter is required to access content hosted on a third-party website.

\begin{table}[htbp]
\small\addtolength{\tabcolsep}{-2pt}
\footnotesize
\begin{tabular}{p{2.9cm}|p{4.9cm}} 
    \hline
    \textbf{Stakeholders} & \textbf{Positioning on cookie walls and paywalls}\\
	\hline
	    EDPB~\cite{WPconsent2020, EDPB-Privacy2018}, EDPS~\cite{EDPS-epriv-2016}, BEUC~\cite{BEUC-ePriv-2017},  Dutch~\cite{DutchDPA-cookiewall}, German~\cite{GermanDPA-cookiewalls-2019}, Danish~\cite{DanishDPA-cookiewall-2020}, Belgian~\cite{BelgianDPA-cookiewall-2020}  DPAs & Cookie walls violate a freely given consent; access to services and functionalities
must not be made conditional on the user consent to non-necessary tracking purposes\\
	\hline
	Italian~\cite{ItalianDPA-cookiewall-2021}, Spanish~\cite{SpanishDPA-cookiewall-2022} DPAS & Cookie walls are not valid. Exceptions: if i)  users are informed, 
ii)	alternative access %
is offered %
without requiring  acceptance to tracking, 
iii) equivalent alternative service %
\\
	\hline
	French DPA~\cite{cnil_cookie_22} & Paywalls  assessed on a case by case analysis according to evaluation criteria: reasonable price, limited to fair remuneration, limited tracking \\
\hline
	    UK DPA~\cite{UKDPA-cookiewalls-2019}  & Unclear position regarding cookie walls. GDPR must be balanced against other rights, including freedom of expression and to conduct a business. \\ \hline
	    Austrian DPA~\cite{datenschutz-behorde_faq_2022} & Paywalls consist of a valid consent    \\ \hline
\end{tabular}
    \caption{Stakeholders positioning regarding *-walls.}
    \label{table:requirements}
    \vspace{-8mm}
\end{table}

\textbf{Lawfulness.}
From the analysis of Table~\ref{table:requirements}, we observe the following: 
\begin{inparaenum}[i)]
    \item several stakeholders present divergent positions regarding cookie walls and paywalls, which hamper the desired level playing field between websites in the EU, as well as legal uncertainty;
    \item regulators tend to accept  paywalls, according to a case by case analysis and evaluation criteria, (\textit{e.g.} price that is "reasonable", or "fair remuneration"). Such casuistic analysis blocks efforts for automated auditing from DPAs and other organizations and researchers to check compliance of website paywalls;
    \item the concept of equivalent service offer needs criteria to assure a balanced offer;
    \item paywalls are dependent on the fact that users are informed of alternative services, though several  user studies report that users do not read or understand the content of consent banners~\cite{Kulyk20206-userreactiontocookiedisclaimer, Borberg2022};
    \item fundamental criteria regarding
    \begin{inparaenum}
    \item the type of data that can be processed, 
    \item the retention period of such data, 
    \item limits on the purposes, 
    \item types of data subjects that might be affected (like children) are  absent in any of these guidelines;
    \end{inparaenum}
    \item news sites might fall in the category of sites revealing information about special categories of data (Article 9(1) GDPR), since reading about certain topics could reveal one's political opinion or health status~\cite{wesselkamp:hal-03241333}.  Also, information about people’s media use and reading habits is generally sensitive~\cite{zuiderveen2017assessment};
    \item the future ePrivacy Regulation proposal~\cite{ePRegulationProposal-Council} (in Recitals 20aaaa and 21aa of the Council's version) already includes the concept of an "equivalent offer" to consent to tracking purposes and thus acknowledges the possibility of paywalls. It prohibits walls by public authorities, and service providers in a dominant position, alongside with the French DPA standpoint. 
\end{inparaenum}

\vspace{-.35cm}
\subsection{The results of the empirical study}
\vspace{-.05cm}

\textbf{Classification.}
Though we initially based our classification on the one developed presented in \cite{papadopoulos_another_2020}, %
it does not accurately represent the current state of %
paywalls because 
i) it does not differentiate soft and metered paywalls,
ii) cookie walls and registration walls are absent of their classification despite their presence on the web,
iii) different communities refer to the same practices by different terms. %
Hence, our classification follows:
\textit{\textbf{Hard paywall}} requires a one-time or a standing payment with money (\textit{i.e.} subscription or enrolment) before any online content can be accessed (as defined by \cite{papadopoulos_keeping_2020}).   
\textbf{\textit{Soft paywall}} presents the beginning of the content to generate interest, but the full content is restricted to payment. 
\textbf{\textit{Metered paywall}} provides users with a certain contingent of articles free of charge that is time bounded. 
\textbf{\textit{Registration wall}} provides users with only one option of creating an account on the website (otherwise users will have denied access).
\textbf{\textit{Cookie wall}} denies users access if they do not consent to all %
    trackers present on %
    that website, regardless of payment.
\textbf{\textit{Cookie paywall}} provides users with two choices: either 1) consent to tracking, and 2)  payment/subscription (by money) to use the website tracking-free.

\textbf{Prevalence of paywall categories.} 
The analysis was conducted on 2800 
websites (see \textbf{Data collection} in Section~\ref{sec:study}):
\begin{inparaenum}[a)]
    \item 61 
    of which contained a paywall (2.72\%  of the analyzed websites),
    \item 13 contain a cookie paywall (0.66\%), wherein 6 thereof had a hard paywall, and
    \item 7 embed a cookie paywall combined with a soft paywall. %
\end{inparaenum}
The results are summarized in Table~\ref{tab:results}.
The percentage of paywalls is approximately three times lower than the percentage found by \cite{papadopoulos_another_2020}, which may indicate a decrease in popularity of this practice: paywalls are not a streamlined practice.
However, this finding must be taken with a pinch of salt, notably because 1) the dataset is limited in terms of scope and size since it only comprises the 150 top websites of central European countries which may bias the general results; but also because 2) we observed that it is complex %
to measure the extent and the scale of this practice with this methodology: manual measurement is intrinsically unpredictable.
Indeed, a second analysis lead to different results.
This second round was conducted using Firefox 102 and Chrome 103 (Chromium flavored) on Ubuntu 20.04 in incognito mode.
We observed discrepancies in real-time between the two browsers, and a general inconsistency of the results according to various parameters, such as the type of browser, the OS, and the IP address.
For instance, on the 26th of July, for the same user using the same machine (OS and IP are therefore identical), the website \url{https://dennikn.sk/} presents at the same time a cookie wall on Chrome and a soft paywall on Firefox (see Figure~\ref{fig:differences} in the Appendix).
We draw two conclusions from these findings: i) the extent of these practices is intrinsically hard to measure, especially with a manual analysis; ii) \textit{web users are presented with different interfaces according to criteria unknown to them}.
An automatic and cross-validated \textbf{large scale analysis} may address this issue: an analysis performed by multiple crawlers (automatic) using different settings (cross-validation), to assess how each parameter may influence the presentation to end-users.
\footnote{Note that a possible technical manner for website to differentiate users is to collect a fingerprint of their web browser.
    Other parameters than the OS and the IP address can then be considered to classify users, such as the list of plugins~\cite{laperdrix_browser_2020}.}
However, even if some reverse engineering may partially explain their working, the lack of transparency from these systems can be seen as deceptive to users.
A large scale analysis can be combined with an \textbf{automated tracking detection}.

\textbf{Types of websites using paywalls.}
A large majority of paywalls, and more specifically cookie paywalls, were found on news media websites which corroborates \cite{papadopoulos_keeping_2020}'s findings. 
One of the arguments in favor of this finding  (\textit{e.g.}, in case of the Austrian DPA) is that \textit{traditional \textit{newspapers and magazines} used to sell their \textit{printed products} to their customers (readers) and this can justify the adoption of paywalls}.~\footnote{The authors do not support this argument.} 
Due to the fact that almost only news and magazine websites used a wall, we had decided to further divide this category into: 
\begin{inparaenum}[i)]
    \item daily newspaper, 
    \item weekly newspaper, 
    \item weekly magazine, 
    \item news magazine,
    \item news portal. 
\end{inparaenum}
The subject-specific magazines that have a cookie wall or paywall in their websites were assigned to the corresponding magazine category, for example, to the category "technology magazine".
Daily newspapers and news portals use paywalls the most - namely a total of 31 daily newspapers and 14 news portals respectively. 
In third place are  weekly magazines wherein 
only four weekly magazines with paywalls were found.

\textbf{Prevalence of TCF framework.}
Our analysis %
confirms the prevalence of the TCF framework (introduced in Section \ref{sec:background}).
Websites using the TCF typically place a cookie named \textit{euconsent-v2} on the user's device once consent has been obtained, and sometimes beforehand.
The content of this cookie can be decoded at \url{http://iabtcf.com/#/decode}, and yields the purposes and vendors for which consent has been obtained (if any).
We found out that 11 out of 13 cookie paywalls are using it.

\textbf{No tracking prior to user consent.} The same tracking assessment indicates compliance with the TCF framework: before any action (besides loading the page), no tracker has been stored (one has been found but on one website which does not use the TCF). Websites using the TCF systematically create a \textit{euconsent-v2} cookie (characteristic of the use of the TCF) \textit{after} %
consent is given to all tracking purposes.
A large scale analysis may lead to different results (see Matte \textit{et al.}~\cite{matte_cookie_2019}). 

\textbf{Prices of paywalls.}
All websites presenting cookie paywalls are subscription-based whereby a subscription buys an unlimited access to all content of the website for one user.
Subscriptions are mostly offered per month (9 websites out of 13), but some propose a per-week subscription.
We observe prices ranging from 2,99€ to 6,25€ per month (after normalization based on the price per year), which amounts to 36 to 75€ per year.
These prices are reasonable compared to the average subscription to a newspaper (the average monthly price for an online newspaper is 14.06€ ~\cite{simon_pay_2019}).
Table~\ref{tab:prices} in the Appendix presents the subscription prices of cookie paywalls normalized per month and per year.

\textbf{Registration wall.} Only one website (www.nzz.ch) was found that only requires registration (without payment) in order to see the full text of any article.

\vspace{-2mm} 

\section{Conclusion and Future work}
\label{sec:future}

Cookie paywalls require more attention from both academia and the legislator: the former to provide more substantial technical and legal analyses, and the latter to level the playing field (see \cite{european_data_protection_board_statement_2022}) towards protective and viable business models across the EU, in line with the current practices and their fast evolution.
Technically, large scale analyses and automated tracking detection would shed light on these opaque practices.
Legally, case-by-case studies, potentially supported by a technical analysis, would facilitate informed decisions by regulators.
Finally, future work includes user-centered studies answering questions such as:
\begin{inparaenum}
    \item Do users actually pay for subscriptions?
    \item Which reasons motivate a possible payment?
\end{inparaenum}
Generally speaking, are cookie paywalls yet another variation of a dark pattern used to nudge users~\cite{humanHumanCentricPerspectiveDigital2021}?

\vspace{-2mm}

\begin{table*}[bp]
\tiny
\caption{Prices of cookie paywalls}
\label{tab:prices}
\hspace*{-.8cm}
\begin{tabular}{@{}lllllllllllllllll@{}}
\toprule
 & DerStandard & Krone & Kurier & Vol & Vienna & Spiegel & Bild & T-online & Welt & Zeit & Heise & Stern & Rp-online &  &  &  \\ \midrule
\begin{tabular}[c]{@{}l@{}}Price in € \\ (normalized over a year)\end{tabular} & 75 & 59.88 & 43.2 & 62.4 & 62.4 & 59.88 & 47.88 & 35.88 & 47.88 & 62.4 & 59.4 & 59.88 & 52 &  &  &  \\
\begin{tabular}[c]{@{}l@{}}Price in € \\ (normalized over a month)\end{tabular} & 6.25 & 4.99 & 3.6 & 5.2 & 5.2 & 4.99 & 3.99 & 2.99 & 3.99 & 5.2 & 4.95 & 4.99 & 4.33 &  &  &  \\
Notes & \begin{tabular}[c]{@{}l@{}}1€/month for 3 months\\ then 8€/month\end{tabular} & 4.99€/month & 3.6€/month & 1,2€/week & 1,2€/week & 4,99€/month & 3,99€/month & 2.99€/month & 3,99€/month & 1,2€/week & 4,95€/month & 4,99€/month & 1€/week &  &  &  \\ \bottomrule
\end{tabular}
\end{table*}

\begin{acks}
\small
This paper has been partially funded by the Internet Foundation Austria (IPA) within the \textit{netidee} call (RESPECTeD-IoT Grant\#5937). 
Cristiana Santos is funded by RENFORCE.
\end{acks}

\bibliographystyle{acm}
\bibliography{bib_noURL}

\appendix

\newpage
\section{Additional Material}
\begin{figure}[!h]
\includegraphics[scale=.25]{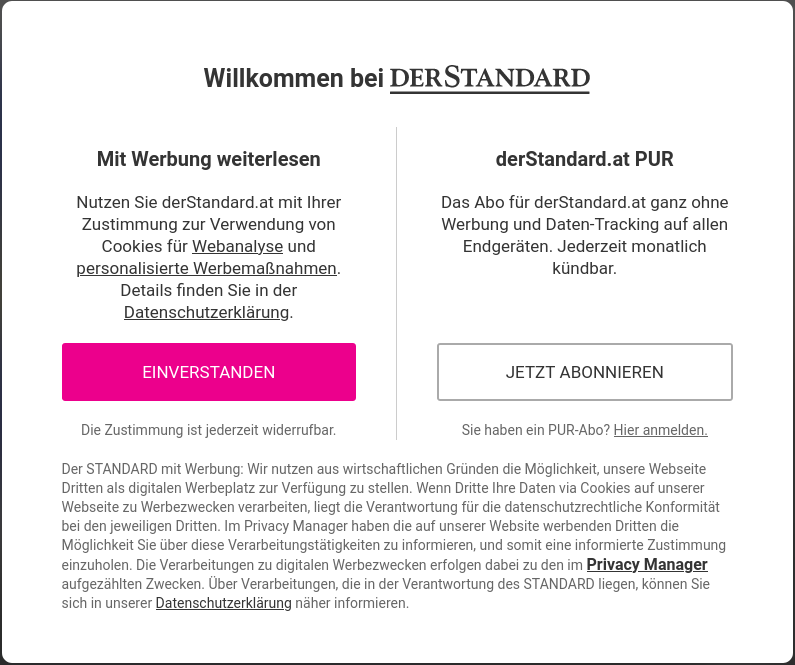}
\caption{Example of a cookie paywall on the DerStandard website (https://www.derstandard.at/consent/tcf/), a leading Austrian newspaper and the highest ranked Austrian paywall. 
The left option proposes to consent to cookies for analytics and personalized advertising.
The literal translation means: "Continue reading with ads. Use derStandard.at with your consent to the use of cookies for web analytics and personalized advertising. Details can be found in the privacy policy. I agree". 
The right option is a monthly subscription free of advertising and tracking. 
Its translation reads "Subscribe to derStandard.at without any advertising or data tracking on all devices. It can be cancelled monthly at any time. Subscribe now."}
\label{fig:derstandard}
\end{figure}

Figure~\ref{fig:derstandard} illustrates a typical example of a cookie paywall. 

\begin{figure}[!hb]
  \begin{subfigure}[b]{0.75\columnwidth}
  \includegraphics[width=\linewidth]{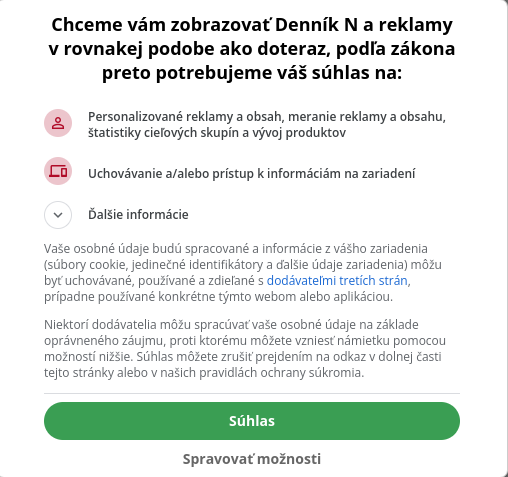}
  \caption{The website Dennikn.sk presents a cookie wall to Chrome users.}
  \end{subfigure}
  \hfill %
  \begin{subfigure}[b]{0.75\columnwidth}
  \includegraphics[width=\linewidth]{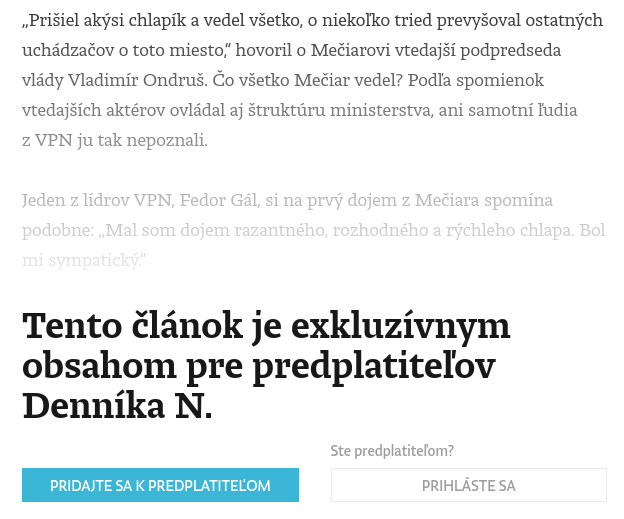}
  \caption{The website Dennikn.sk presents a soft paywall to Firefox users.}
  \end{subfigure}
  \caption{The same website presents different types of walls to the same user according to her web browser.}
\label{fig:differences}
\end{figure}

Figure~\ref{fig:differences} illustrates that for the same user using the same machine (OS and IP are therefore identical), the website \url{https://dennikn.sk/} presents at the same time a cookie wall on Chrome and a soft paywall on Firefox: we observed discrepancies in real-time between two browsers according to various parameters such as the type of browser, the OS, and the IP address.

\end{document}